\documentstyle[12pt]{article}
\begin{document}
\title{Second Virial Coefficient For Real Gases \\ At High
Temperature\thanks{This work is supported in part by funds provided by
the U.S. Department of Energy (D.O.E.) under cooperative research
agreement \#DF-FC02-94ER40818 and Conselho Nacional de Desenvolvimento
Cient\' \i fico e Tecnol\'ogico (CNPq) - Brazilian agency.}}
\author {{\rm H. Boschi-Filho\thanks{E-mail address:
boschi@ctp.mit.edu} $^\ddagger$ and C.C.Buthers$^\ddagger$}\\ \it
$^\dagger$ Center for Theoretical Physics, Laboratory for Nuclear
Science\\ \it Massachusetts Institute of Technology\\ \it Cambridge,
Massachusetts 02139-4307, USA\\ and \\ \it$^\ddagger$Instituto de
F\'\i sica, Universidade Federal do Rio de Janeiro \\ \it Cidade
Universit\'aria, Ilha do Fund\~ao, Caixa Postal 68528 \\ \it 21945-970
Rio de Janeiro, BRAZIL}
\date{January 24, 1997}
\maketitle
\begin{abstract} We study the second virial coefficient, $B(T)$, for
simple real gases at high temperature. Theoretical arguments imply
that there exists a certain temperature, $T_i$, for each gas, for
which this coefficient is a maximum. However, the experimental data
clearly exhibits this maximum only for the Helium gas. We argue that
this is so because few experimental data are known in the region where
these maxima should appear for other gases. We make different
assumptions to estimate $T_i$. First, we adopt an empirical formulae
for $B(T)$. Secondly, we assume that the intermolecular potential is
the Lennard-Jones one and later we interpolate the known experimental
data of $B(T)$ for {\it Ar, He, Kr, $H_2$, $N_2$, $O_2$, Ne} and $Xe$
with simple polynomials of arbitrary powers, combined or not with
exponentials. With this assumptions we estimate the values of $T_i$ for
these gases and compare them.
\end{abstract} \vfill \par
 \par

\pagebreak
\baselineskip = 20pt

\begin{section}{Introduction} \setcounter{equation}{0}

Long time ago, Kamerlingh Onnes and collaborators performed a
systematical study of physical properties of gases at low temperature,
measuring their deviations from the ideal gas law, or in other words,
determining their virial coefficients \cite{KB}-\cite{KU}.  Since
then, much research has been done in this direction collecting more
and more data for many different substances, mainly at low temperature
\cite{DS}. In particular, in 1925, Holborn and Otto studied some
simple gases (Ar, He, H$_2$, N$_2$, and Ne) and they were able to
determine an empirical formula for the second virial coefficient,
$B(T)$, which constituted at that time and persists until today as a
land-mark work in physics \cite{HO}.

On the other hand and at the same time, Lennard-Jones showed in
another remarkable work that for intermolecular potentials written as
binomials of arbitrary powers of the intermolecular distance, the
corresponding second virial coefficient $B(T)$ can be calculated
exactly as an infinite power series \cite{JL}. This kind of potential
describes very well the behavior of simple molecules as the one
studied by Kamerlingh Onnes, Holborn and Otto and others. Also, other
types of intermolecular potentials have been discussed but without
exact integrability \cite{Hisch}. Despite of the long time that these
empirical and theoretical expressions for $B(T)$ are known, they are
usually not equal. This can be understood paying attention to the fact
that the experimental data available for most substances are
restricted to a certain range of temperatures, usually below $600
K$. The choice for low temperature physics seems rather natural if
one, for example, is looking for critical phenomena in condensed
matter which can be almost entirely found in this region. However, the
high temperature behavior of real gases can also be of interest, for
example, in the study of hot plasma and in many situations in
astrophysics as nucleosynthesis, super-novae, etc.

Here in this paper, we are going to examine the second virial
coefficient of some real simple gases (Ar, He, Kr, H$_2$, N$_2$,
O$_2$, Ne and Xe) at high temperature. This is not an easy task since
few data for these and other gases are available for temperatures
above the range $600 \sim 1000 K$. The main point of this work is that
we can show from measures of $B(T)$ that most of these real simple
gases admits a maximum value corresponding to a certain temperature
for each gas, which may be called an inversion temperature
($T_i$). The existence of this maximum may be not a surprise once it
is already present, for the Helium, in Holborn and Otto work
\cite{HO}. However, for all other real gases they are not yet
known. Anyway, some previous discussions on this maximum \cite{MA} and
on this inversion temperature \cite{RJ} can be found in the
literature.

This paper is organized as follows: in section 2, we discuss the
inversion temperature $T_i$ associated with the Joule or free
expansion and recall the well known inversion temperature related to
the Joule-Thomson throttling process, $T_{iJT}$. We show that there is
a kind of duality between the equations governing these two
temperatures. The temperatures $T_{iJT}$ are known experimentally and
can also be obtained by various theoretical methods, one of them
assumes the knowledge of an equation of state. We show, however, that
the known equations of state do not lead to any finite $T_i$. In
section 3, we employ different methods to estimate the inversion
temperature $T_i$. The first and simplest one is to use the empirical
expression for $B(T)$ given by Holborn and Otto from which we can find
easily its maximum.  The second and third methods are based on the
assumption that the intermolecular potential between the molecules of
the real gas is the Lennard-Jones 6-12 potential \cite{JL}. So, the
second method consists in finding the inversion temperature $T_i$
evaluating $dB/dT = 0$ numerically. The third method is a variation of
the second since the assumption of the Lennard-Jones potential permit
us to relate the temperatures $T_i$ and $T_{iJT}$, determining the
former, once the latter are well known. The fourth and last method for
determining $T_i$, that we discuss in this paper, is a numerical
analysis of the experimental data for $B(T)$ for the above mentioned
gases. We interpolate these data with simple polynomial functions with
arbitrary powers with and without exponentials terms. In section 4, we
compare the results obtained from these functions with the ones from
other methods and present our conclusions.

\end{section}

\begin{section}{The Inversion Temperatures $T_i$ and $T_{iJT}$} 
\setcounter{equation}{0}

The inversion temperature $T_i$, for which $B(T)$ is a maximum, has a
simple physical interpretation. A real gas, initially at this
temperature, subjected to a free expansion remains at this
temperature for any change in its volume, which is the ideal gas
behavior, for any initial temperature. If the real gas initial
temperature is greater than $T_i$ then it will get hotter in a free
expansion and the reverse occurs if $T < T_i$, which is our "daily"
experience. One can understand this simply by noting that the Joule
coefficient, i.e., the variation of temperature against volume in a
free expansion (with internal energy $U$ being constant) can be
expressed as \cite{Cast}

\begin{equation}
\label{Joule}
J\equiv {({\partial T\over
\partial V})}_U={1\over c_V}[P - T({\partial P\over \partial
T})_V].
\end{equation} 

\noindent Writing the virial expansion as

\begin{equation}
\label{virial}
PV=RT(1+{B\over V}+{C\over V^2}+{D\over V^3}+{E\over V^4}+...)
\end{equation}

\noindent it is easy to show that, in the thermodynamical limit ($V
\rightarrow \infty$)

\begin{equation}
\label{Joule2}
J = -{RT^2\over c_VV^2}({dB\over dT}),
\end{equation}

\noindent where $B \equiv B(T)$. Looking at the above expression one
can see that the maximum of $B(T)$, which may occur for certain values
of $T = T_i$, vanishes the Joule coefficient $J$:

\begin{equation}
\label{Joule3}
J = 0\hskip 1cm{\Leftrightarrow}\hskip 1cm{{dB\over {dT}} = 0}.                
\end{equation}

\noindent Assuming that $T_i$ is unique, above this temperature $J$
will be positive and so the temperature increases under a free
expansion. Bellow $T_i$ the temperature decreases.

In fact, this behavior is also expected in the bases of the Yang and
Lee Theorem \cite{YL} (see also \cite{MA}) if one assumes that the
intermolecular potential of the gas has a shape as the one given by
fig.1. This is a typical shape which includes the well known
Lennard-Jones, Stockmayer and other potentials \cite{Hisch}. This kind
of potential implies a maximum for $B(T)$ since they are not infinite
for finite distances between molecules, contrary to what happens for
the hard-sphere case, for example.

At this point, it is interesting to discuss a well known inversion
temperature, $T_{iJT}$, that one associated with the Joule-Thomson
throttling process with constant enthalpy and its relation to the
previous discussed inversion temperature $T_i$. The relevant
coefficient in this case can be written as \cite{Cast}
\begin{equation}
\label{JT}
\mu={1\over c_P}[T({\partial V\over \partial T})_P -V].
\end{equation}

\noindent As the derivative of the volume in respect to the
temperature must be calculated under constant pressure, the virial
expansion (\ref{virial}) is not the most appropriate in this case and
it is usual to rewrite it as
\begin{equation}
\label{pressao}
PV=RT(1+B_PP+C_PP^2+D_PP^3+E_PP^4+...),
\end{equation}

\noindent so we get immediately, in the low pressure limit ($P
\rightarrow 0$)

\begin{equation} \label{mi} \mu = {RT^2\over c_P}({dB_P\over dT}).
\end{equation} 

\noindent
The inversion temperature, in this process, $T_{iJT}$,
is determined by the vanishing $\mu$: 
\begin{equation} \label{mi3} \mu
= 0 \hskip 1cm{\Leftrightarrow}\hskip 1cm{{dB_P\over
{dT}}\vert_{T=T_{iJT}}=0}.  \end{equation}

\noindent These temperatures, apart from being well known for most
real gases, are of great practical importance, for example, to improve
the efficiency of thermal engines as refrigerators \cite{An}. Another
striking property of this temperature is its resemblance with
$T_i$. This is not a coincidence since the second virial coefficients
at constant volume and pressure, $B$ and $B_P$, respectively, are
simply related by
\begin{equation}
B=B_PRT.
\end{equation}
\noindent 
So, we can trace a parallel between these two inversion temperatures
first substituting the above expression into eq.(\ref{mi}) so we get
\begin{equation} \label{mi2} 
\mu=0 \hskip 1cm{\Leftrightarrow} \hskip
1cm{{dB\over {dT}} = {B\over T};}\hskip 1cm{T=T_{iJT},} \end{equation}

\noindent which is the usual expression for the
throttling process \cite{Hisch} and then into eq. (\ref{Joule2})
\begin{equation}
\label{Joule4}
J=0 \hskip 1cm{\Leftrightarrow}\hskip 1cm{{dB_P\over {dT}} = -{B_P\over
T};\hskip 1cm{T=T_i},}
\end{equation}

\noindent for the Joule expansion. Comparing eqs. (\ref{Joule3}) and
(\ref{mi3}), (\ref{mi2}) and (\ref{Joule4}) one can see clearly a kind
of duality between the equations governing these two processes: The
role played by the inversion temperature $T_{iJT}$ in the constant
enthalpy process is analogous to the $T_i$ in the constant energy
case. Despite of this appealing resemblance the inversion temperature
$T_i$ for the Joule process has been rarely discussed \cite{RJ}, and
these discussions are for from being satisfactory or complete.

The temperatures $T_{iJT}$ are known experimentally and can also be
estimated using simple equations of state \cite{Cast}. One can wonder
if the inversion temperature $T_i$ can be found from equations of
state as happens for $T_{iJT}$. This should be nice, but for various
equations of state the corresponding second virial coefficient does
not allow any finite $T_i$, as can be seen by inspection of Table 1.
Note that $B(T)$ for the Beattie-Bridgeman equation of state admits in
principle a maximum, but actually as the values of the constants $a$
and $c$ are positive for real gases one finds that no real $T_i$ are
admissible from this equation too.

\end{section}

\begin{section}{Estimates On The Inversion Temperature $T_i$}
\setcounter{equation}{0}

There are various possible ways of computing $T_i$. We will discuss
some of them and compare their results. If one knows the correct
expression for $B(T)$ then the problem would be trivial, since the
equation $dB/dT=0$ could be solved immediately. But what we have
until now are, on one side expressions for $B(T)$ based on theoretical
assumptions and on the other side empirical expressions which differ
in general from each other and are constructed on the data for which
the temperatures are low, usually under $600K$. As we are going to
show, these maxima appear in a region beyond this temperature, except
for the Helium, for which $T_i\simeq 200 K$ \cite{HO}. 

Among the known expressions for $B(T)$, the most simple is, perhaps,
the Holborn and Otto \cite{HO} one

\begin{equation}
\label{HO}
B(T)=a+bT+{c\over T}+{e\over T^3},
\end{equation}

\noindent from which one can easily find that the inversion
temperature, in this case is given by
\begin{equation}
\label{HO2}
T_i=\sqrt{c-\sqrt{c^2+12eb}\over 2b}.
\end{equation}

\noindent The values of the coefficients appearing in eq. (\ref{HO}) and
the corresponding $T_i$'s for some simple gases are found in Table 2.

From statistical mechanics, it is well known that an intermolecular
potential, $u(\vec{r})$, and the corresponding the second virial
coefficient are related by:
\begin{equation}
\label{stat}
B(T)\;=\;{1\over 2}\int_{V} d\vec{r} \; [1-\exp( - u(\vec{r})/kT)],
\end{equation}

\noindent where $k$ is the Boltzmann constant. So, the inversion
temperature $T_i$ could also, in principle, be obtained analytically
through this expression. However, this is a formidable task for almost
of known intermolecular potentials except from the trivial ones like
the hard-sphere, which does not imply any finite $T_i$. An important
exception is the Lennard-Jones potential \cite{JL}:
\begin{equation}
\label{LJ}
u_{LJ}(r) = \epsilon_0[({r_0\over r})^{12} - 2({r_0\over r})^6],
\end{equation}

\noindent for which $B(T)$ can be computed exactly as an infinite
power series. The constants $\epsilon_0$ and $r_0$ are tabulated for
many real gases. Substituting (\ref{LJ}) into (\ref{stat}) one can show
that \cite{Hisch}
\begin{equation}
\label{soma}
B(T)=b_0\sum_{j=0}^\infty {b^{(j)}T^{*^{-{(2j+1)\over4}}}},
\end{equation}

\noindent where $T^{*}={kT\over \epsilon_0}$ is the reduced
temperature, $b_0 ={\sqrt{2}\over3}\pi\tilde{n}r_0^3$, $\tilde{n}$
being the number of molecules per mol and
\begin{equation}
b^{(j)}=-{2^{j+{1\over 2}}\over {4j!}}\Gamma({2j-1\over 4}).
\end{equation}

\noindent The importance of this potential is related to the fact that
it describes the Van der Walls force of attraction between molecules,
which is proportional to $r^{-7}$(in the non-relativistic limit) and
includes a finite repulsion at small finites distances, being
integrable and fitting (low temperature) data quite well for a great
number of real gases. For this potential, we can evaluate the
inversion temperature $T_i$, since
\begin{equation}
\label{db}
{dB(T)\over{dT}} = - {\epsilon_0 b_0\over{kT^{*}}}\sum_{j=0}^\infty
{{(2j+1)\over 4}b^{(j)}T^{*^{-{(2j+1)\over4}}}}
\end{equation}

\noindent and imposing that
\begin{equation}
{dB(T)\over{dT}} = 0,
\end{equation}

\noindent which solution can be evaluated numerically. As the expected
inversion temperature is of order $10^2 \sim 10^3K$, we see that the
series in eq.(\ref{db}) is rapidly convergent. Taking $j=5$ we find
\begin{equation}
\label{Temp1}
T_i^{*}={kT_i\over \epsilon_0}\cong 25.152
\end{equation}

\noindent with precision of 0.001. If we extend this calculation until
$j=15$, for example, it will change less than 0.001 (see Table
3). Using the data for $\epsilon_0$ from \cite{Hisch} in the above
equation, we can estimate the inversion temperature for mono-atomic and
diatomic (and perhaps for simple poly-atomic gases) which results can
be seen in Table 4.

A related analysis can also be made for the inversion temperature
$T_{iJT}$ of the throttling process. Starting from the expression for
$B(T)$, eq. (\ref{soma}), corresponding to the Lennard-Jones potential
and imposing the conditions (\ref{mi2}) we find (see Table 3)
\begin{equation}
\label{Temp2}
T_{iJT}^* = {k\over \epsilon_0}T_{iJT} \cong 6.431
\end{equation}

As these temperatures are well known, comparing eqs.(\ref{Temp1}) and
(\ref{Temp2}) we can also find $T_i$ from them:
\begin{equation}
\label{rel}
T_i \; \simeq \;3.911 \; T_{iJT}\;,
\end{equation}

\noindent when we assume that the intermolecular potential is the
Lennard-Jones one. The temperatures calculated using this relation are
shown in Table 5. These numerical results could also be inferred from
tabulated values of $B(T)$ and its derivatives for the Lennard-Jones
potential \cite{Hisch} (see also \cite{RJ}).

As our last method for computing $T_i$, we will consider polynomial
expressions for $B(T)$, combined or not with exponential terms, which
parameters will be fixed by best fitting of experimental data from
\cite{DS}. The first and simplest case is
\begin{equation}
\label{b1}
B_1(T)={a\over T^{b}} + {c\over T^{d}}
\end{equation}

\noindent and we will use Powell's method from {\it Numerical
Recipes} \cite{recipes} to minimize $\chi^2$, defined by
\begin{equation}
\chi^2=\sum_{n=1}^N[B_L(T_n)-B_{exp}(T_n)]^{2}
\end{equation}

\noindent where $B_L(T_n)$, with $L=1$, is the expression (\ref{b1})
calculated with $N$ experimental temperature values $T_n$,
$(n=1,...,N)$ and $B_{exp}(T_n)$ are the corresponding experimental
values for the second virial coefficient. In Table 6, we give the
coefficients of (\ref{b1}) which minimize $\chi^2$ for each gas and
present the inversion temperature $T_i$ calculated from them. In fact,
except for the Helium, these values of $T_i$ correspond to
extrapolations on the interpolated expressions of $B_1(T)$.

In order to improve these results we also consider other functions for
$B(T)$, which are
\begin{equation}
\label{b2}
B_2(T)={a\over T^{b}} + {c\over T^{d}}+{e\over T^f}
\end{equation}
\begin{equation}
\label{b3}
B_3(T)={a\over T^{b}} + {c\over T^{d}} + f\exp(-gT)
\end{equation}
\begin{equation}
\label{b4}
B_4(T)={a\over T^b} + {c\over T^d} + {e\over T^f} + g\exp(-hT).
\end{equation}

\noindent and repeat the above procedure, as was done for
$B_1(T)$. The corresponding results for $B_2(T)$, $B_3(T)$ and
$B_4(T)$ are shown in Tables 7, 8 and 9, respectively. The data we
have used for $B(T)$ for these gases are the ones which were compiled
by Dymond and Smith {\cite{DS}}. In particular, for the Argon we used
$N =$ 36 experimental data from \cite{HO}, \cite{TM}, \cite{WL} and
\cite{FH}. For the Helium we used $N =$ 31 experimental data from
\cite{HO},\cite{WR}, \cite{WM} and \cite{KM}.  For the Krypton we used
$N =$ 28 experimental data from \cite{BB}, \cite{WS} and
\cite{RS}. For the Hydrogen we used $N =$ 30 experimental data from
\cite{KB}, \cite{HO}, \cite{TM} and {\cite{GM}}. For the Nitrogen we
used $N =$ 38 experimental data from \cite{KU}, \cite{HO}, \cite{WM}
and \cite{HR}. For the Neon we used $N = $~19 experimental data from
\cite{HO} and \cite{NS}. For the Oxygen we used $N =$ 20 experimental
data from  \cite{KK}, \cite{HO}, \cite{MSG} and \cite{NK}. Finally, for
the Xenon we used $N =$ 30 experimental data from \cite{BBB},
\cite{MWL} and \cite{WLS}.

To get more confidence on these results, we have also shown them
graphically in figures 2 to 9. In these figures we have plotted the
best fitting for eqs. (\ref{b1}) and (\ref{b2})  - ({\ref{b4}}), for each gas
separately. Note that for the most of gases only one curve can be seen
since the fittings are very close to each other. In particular, in
fig. 3 (Helium) three different curves can be seen, corresponding to
the equations (\ref{b1}) and (\ref{b2})  - ({\ref{b4}}). In this case the fitting
represented by eqs. (\ref{b3}) and (\ref{b4}) are very close to each
other.
 
\end{section}

\begin{section}{Discussion and Conclusions}

We have estimated the inversion temperature $T_i$ related to the Joule
or free expansion by different methods. As can be seen from tables 2
and 4 to 9, each gas has a different inversion temperature $T_i$,
which depend also on the method employed. In some cases as for the
Hydrogen the values found can differ by a factor of 2 and for the
Nitrogen by 4. Another gases as the Helium and Argon the discrepancy
was about a factor of 5/3 and 3/2, respectively. Taking this into
account and using a $\chi^2$-weighted average, we estimate a value for
$T_i$ corresponding to each gas, basically from the numerical fitting
expressed in Tables 6~--~9, which results can be seen in Table 10. The
choice for taking only the temperatures from Tables 6~--~9 is based on
the fact that the values from these tables are the ones in which
minimum modeling assumptions were made, since we left the powers of
the polynomials (\ref{b1}) and (\ref{b2}) - ({\ref{b4}}) to be fixed
by best fitting to experimental data, contrary to what happens in the
other methods that we discussed before. Looking at Table 10 one can
also see that for the Oxygen and Krypton we could not find values for
$T_i$, despite that from Tables 4 and 5 we can see, by other methods
some values for them.

To give an idea of the global behavior of the expressions for $B(T)$
and to visualize some of the maxima corresponding to the inversion
temperature $T_i$ discussed above, we have plotted in a single graph
(see fig. 10) the shape of eq. ({\ref{b4}}) which we found for all the
gases discussed in this paper until temperatures of the order of $4
\times 10^3 K$. As one can see from this graph some gases as the
Helium, of course, and the Hydrogen exhibits clearly these maxima. For
some others as the Neon, Argon and Nitrogen these maxima can be seen
in a careful analysis, while for the Krypton, Xenon and Oxygen they
can not be seen in this range of temperatures.

To conclude, we can say that the inversion temperature $T_i$
associated with the free expansion of a real gas can be determined for
some simple gases and in general they were not known because there are
few experimental data for the range of temperatures in which they
should appear. Despite that the physical interpretation of the
inversion temperature $T_i$ is related to a free expansion which is a
non-equilibrium process and so difficult to analyze directly, the
values for $B(T)$ can be taken from usual $PVT$ measures. An important
consequence of determining precisely this inversion temperature is
that it will permit a greater confidence between the experimental and
theoretical expressions for $B(T)$ and consequently on the force
between molecules.

\bigskip
\bigskip
\noindent {\bf Acknowledgments:} We would like to acknowledge
F.M.L. de Almeida, A.S. de Castro and Y.A. Coutinho for the help with
different computational parts of this work and specially A. Ramalho
who wrote a Fortran program which enable us to use the Powell routine
of {\it Numerical Recipes}. H.B.-F. acknowlegdes R. Jackiw for his
hospitality at MIT and for reading the manuscript and K. Huang for an
interesting discussion on these results. The authors were partially
supported by CNPq -- Brazilian agency (C.C.B. under the CNPq/PIBIC
program).

\end{section}

\vfill
\par
\pagebreak

\centerline{\bf Figure Captions}
\bigskip
\bigskip
\bigskip
\bigskip
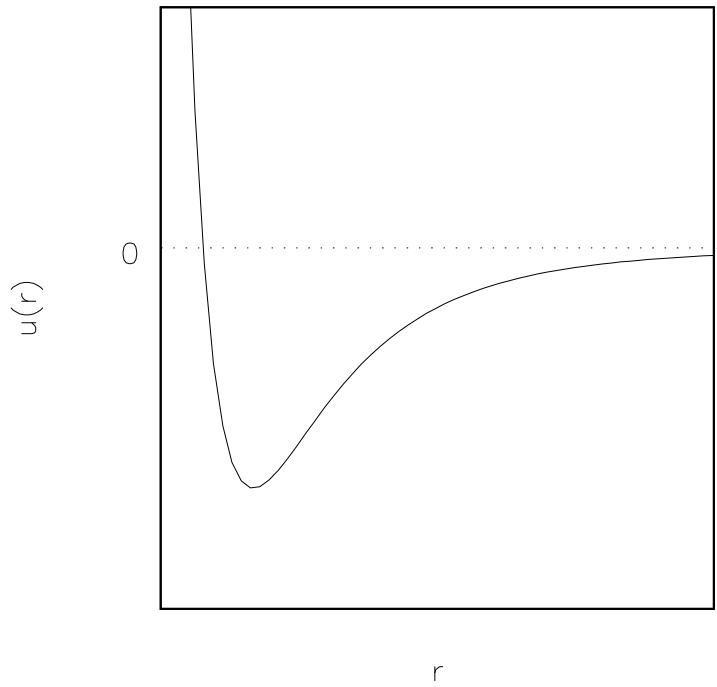
\noindent {\bf Fig. 1:} A typical shape for intermolecular potential
as the one given by the Lennard-Jones potential.
\bigskip\par

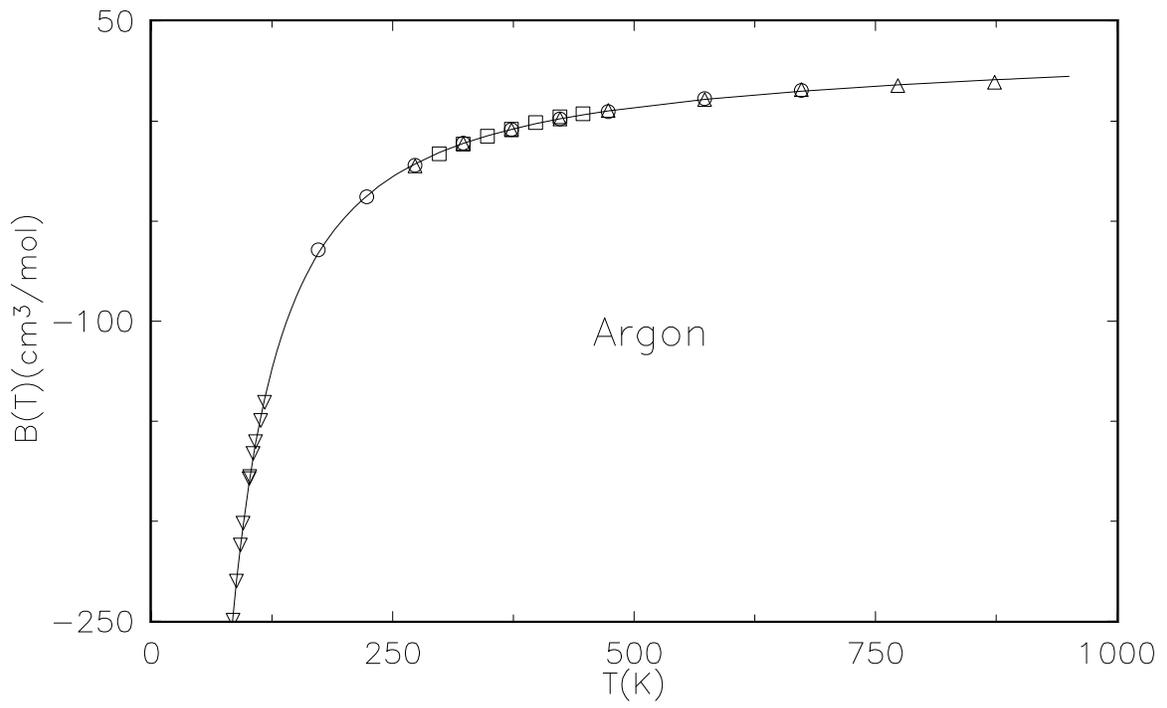
\noindent {\bf Fig. 2:} The second virial coefficient, $B(T)$, for the
Argon.  The curve represents eqs.(\ref{b1}) and (\ref{b2}) -
(\ref{b4}) best fitting experimental data from: $\bigcirc$ -- Holborn
and Otto \cite{HO}; $\sqcup\!\!  \!\!\sqcap$ -- Tanner and Masson
\cite{TM}; $\bigtriangleup$ -- Walley, Lupien and Schneider \cite{WL};
$\bigtriangledown$ -- Fender and Halsey \cite{FH}.
\bigskip\par

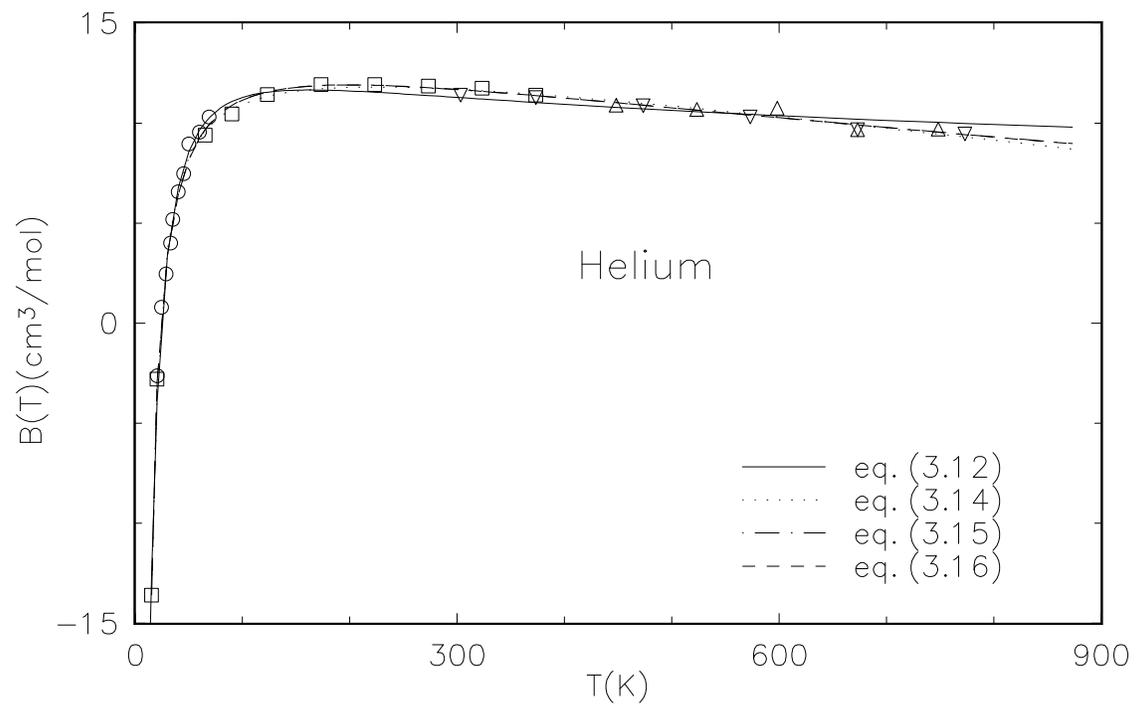
\noindent {\bf Fig. 3:} The second virial coefficient, $B(T)$, for the
Helium.  The curves represent eqs.(\ref{b1}) and (\ref{b2}) -
(\ref{b4}) best fitting experimental data from: $\sqcup\!\!\!\!\sqcap$
-- Holborn and Otto \cite{HO}; $\bigcirc$~--~White, Rubin, Camky and
Johnston \cite{WR}; $\bigtriangleup$~--~Witonsky and Miller \cite{WM};
$\bigtriangledown$ -- Kalfoglou and Miller \cite{KM}. Note that
eqs. (ref{b3}) and (\ref{b4}) can not be distinguisehd in this graph. 
\bigskip\par

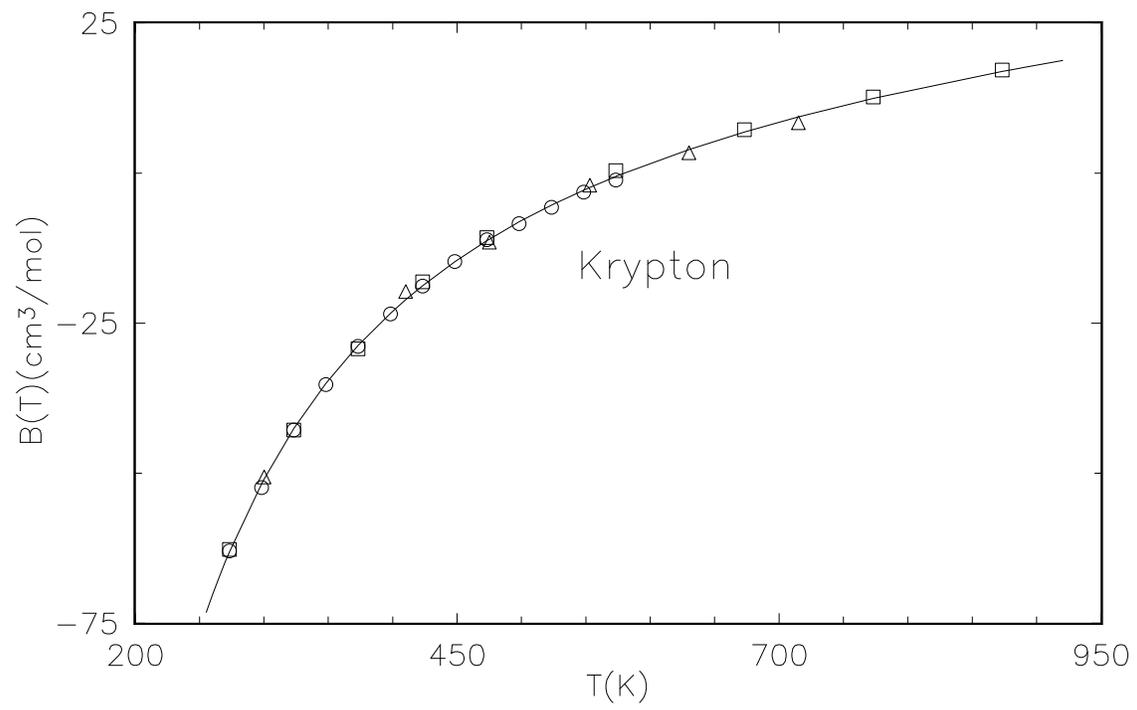
\noindent {\bf Fig. 4:} The second virial coefficient, $B(T)$, for the
Krypton.  The curve represents eqs.(\ref{b1}) and (\ref{b2}) -
(\ref{b4}) best fitting experimental data from: $\bigcirc$ -- Beattie,
Brierley and Barriaut~\cite{BB}; $\sqcup\!\! \!\!\sqcap$ -- Walley and
Schneider \cite{WS}; $\bigtriangleup$ -- Rentschler and Schramm
\cite{RS}.
\bigskip\par

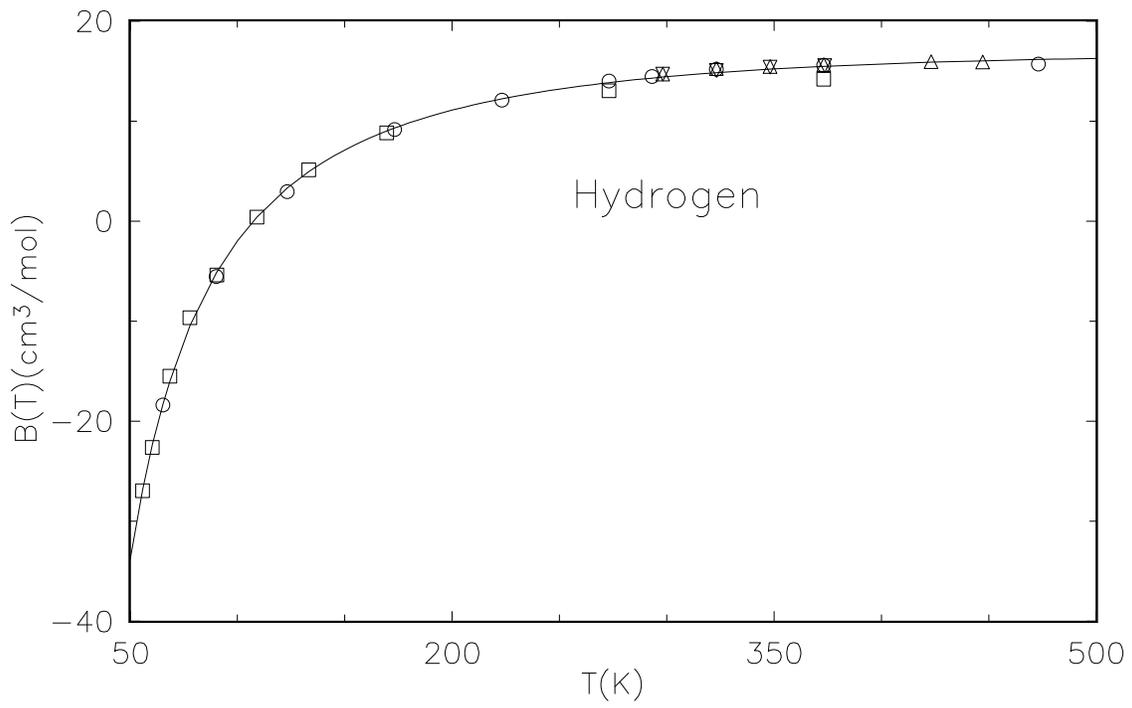
\noindent {\bf Fig. 5:} The second virial coefficient, $B(T)$, for the
Hydrogen. The curve represents eqs.(\ref{b1}) and (\ref{b2}) -
(\ref{b4}) best fitting experimental data from: $\sqcup\!\!
\!\!\sqcap$ -- Kamerlingh Onnes and Braak \cite{KB}; $\bigcirc$ --
Holborn and Otto \cite{HO}; $\bigtriangleup$ -- Tanner and Masson
\cite{TM}: $\bigtriangledown$ -- Gibby, Tanner and Masson~\cite{GM}.
\bigskip\par

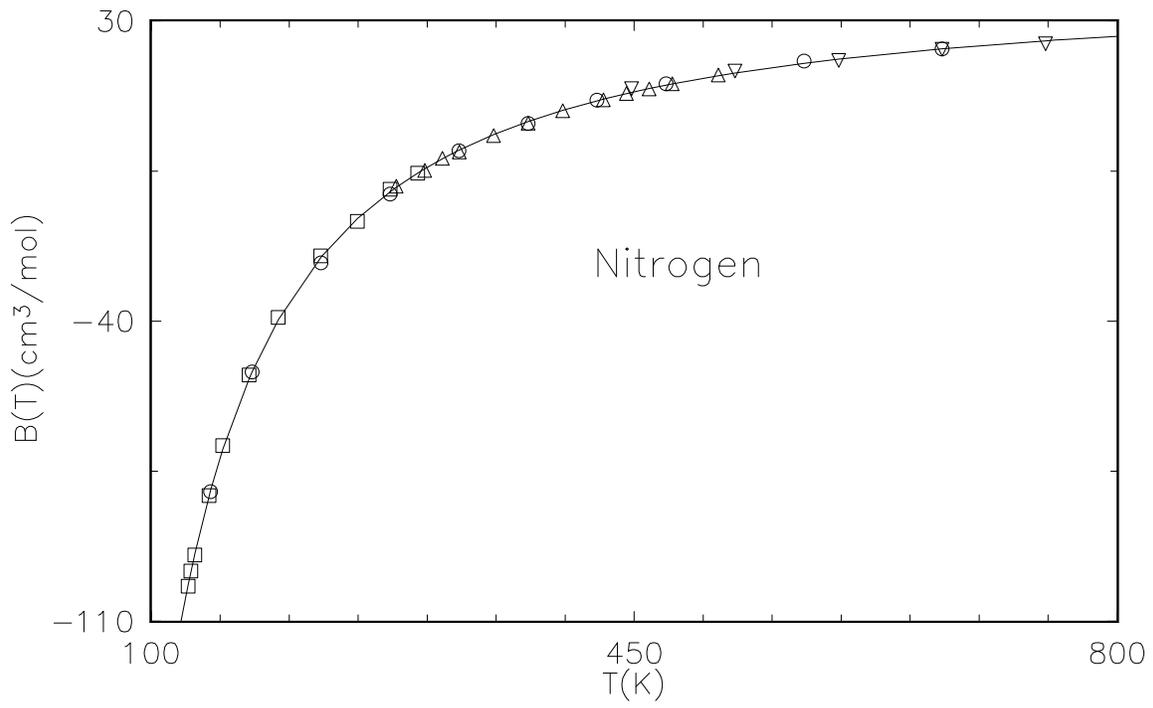
\noindent {\bf Fig. 6:} The second virial coefficient, $B(T)$, for the
Nitrogen. The curve represents eqs.(\ref{b1}) and (\ref{b2}) -
(\ref{b4}) best fitting experimental data from: $\sqcup\!\!
\!\!\sqcap$ -- Kamerlingh Onnes and Van Urk \cite{KU};
$\bigcirc$~--~Holborn and Otto \cite{HO}; $\bigtriangledown$ --
Witonsky and Miller \cite{WM}; $\bigtriangleup$ -- Huff and Reed
\cite{HR}.
\bigskip\par

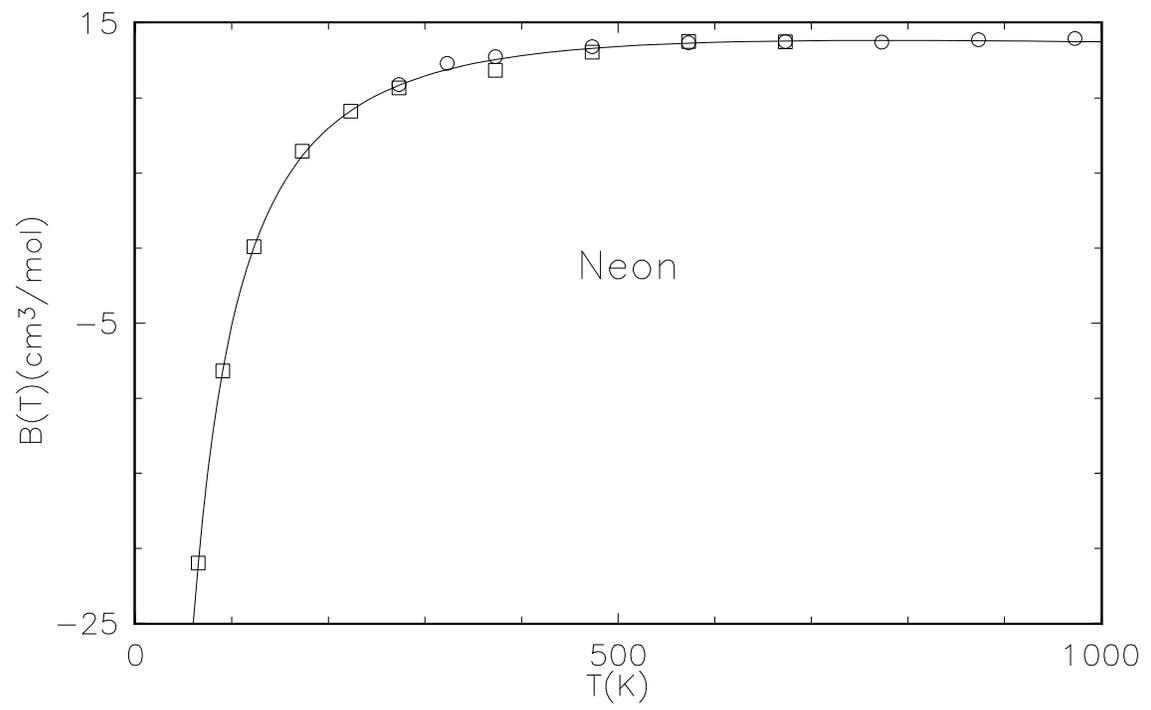
\noindent {\bf Fig. 7:} The second virial coefficient, $B(T)$, for the
Neon.  The curve represents eqs.(\ref{b1}) and (\ref{b2}) - (\ref{b4})
best fitting experimental data from: $\sqcup\!\! \!\!\sqcap$ --
Holborn and Otto \cite{HO}; $\bigcirc$ -- Nicholson and Schneider
\cite{NS}.
\bigskip\par

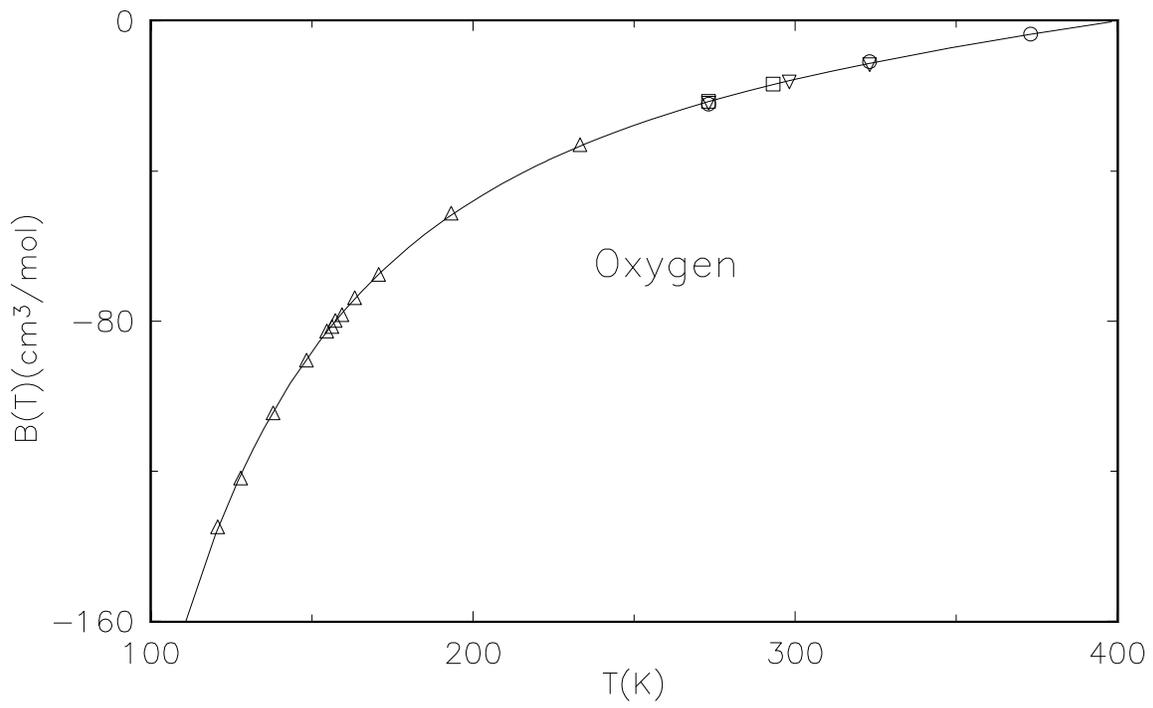
\noindent {\bf Fig. 8:} The second virial coefficient, $B(T)$, for the
Oxygen.  The curve represents eqs.(\ref{b1}) and (\ref{b2}) -
(\ref{b4}) best fitting experimental data from: $\sqcup\!\!
\!\!\sqcap$ -- Kuypers and H. Kamerlingh Onnes \cite{KK}; $\bigcirc$
-- Holborn and Otto \cite{HO}; $\bigtriangledown$ -- Michels, Schamp
and Graaff \cite{MSG}; $\bigtriangleup$ -- Nijhoff and W.H. Keesom
\cite{NK}.
\bigskip\par

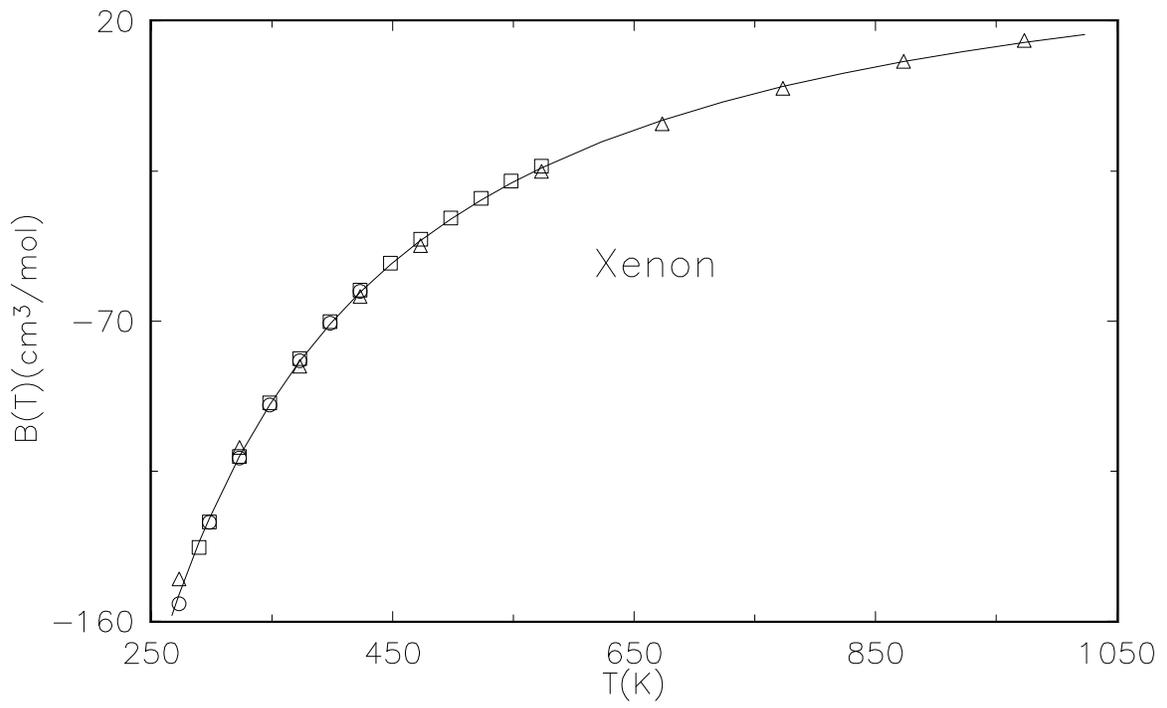
\noindent {\bf Fig. 9:} The second virial coefficient, $B(T)$, for the
Xenon. The curve represents eqs.(\ref{b1}) and (\ref{b2}) - (\ref{b4})
best fitting experimental data from: $\sqcup\!\! \!\!\sqcap$ --
Beattie, Barriault and Brierley \cite{BBB}; $\bigcirc$ -- Michels,
Wassenaar and Louwerse \cite{MWL}; $\bigtriangleup$ -- Walley, Lupien
and Schneider \cite{WLS}.
\bigskip\par

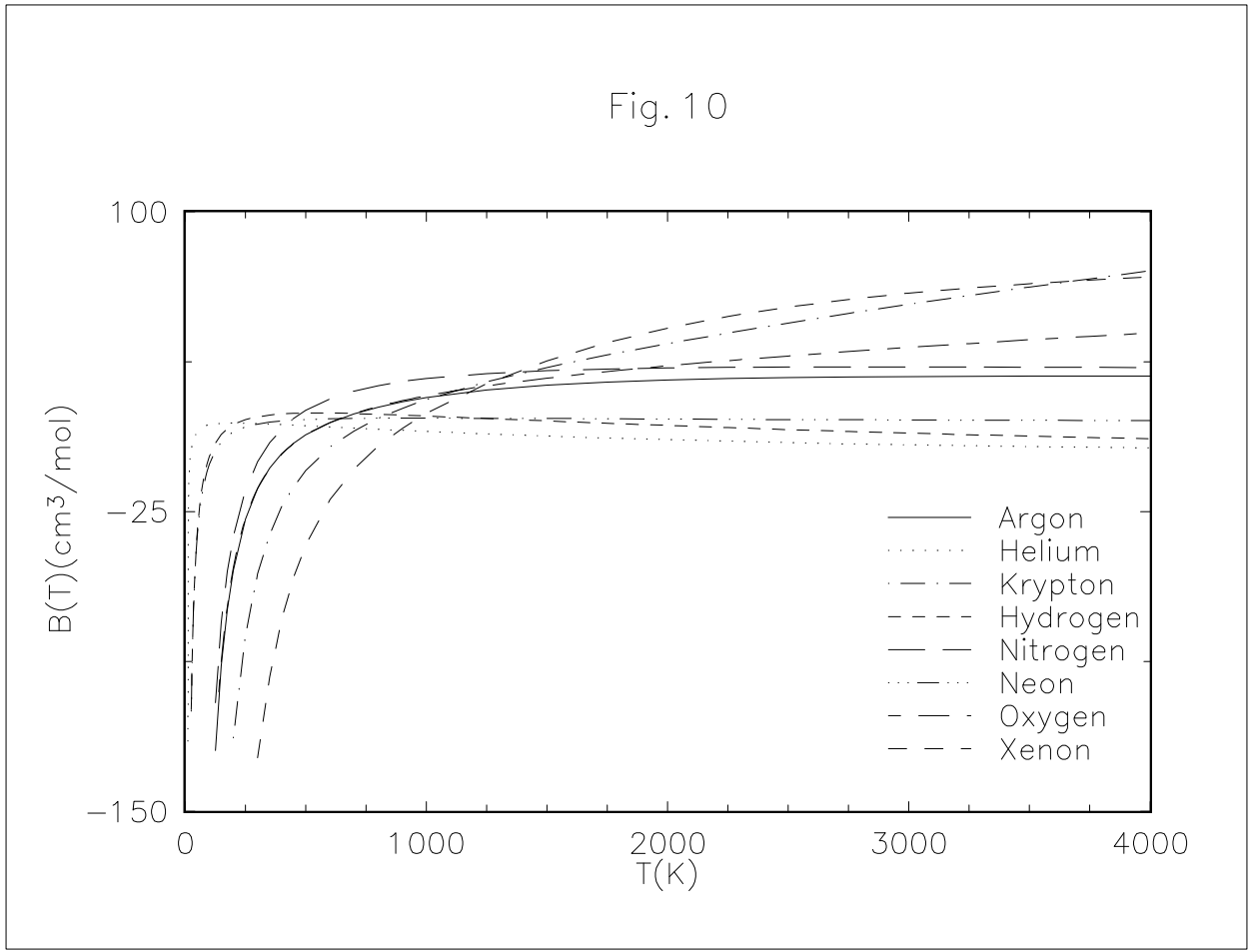
\noindent {\bf Fig. 10:} An overview of the second virial coefficients
represented by eq.(\ref{b4}) as the best fit for the gases discussed
in this paper. We extrapolated these functions to high temperatures in
order to search for their maxima. As one can see, some of them clearly
exhibits a maximum, as the case of Helium and Hydrogen; for Neon,
Argon and Nitrogen they appear slightly and for the others we can not
see them, at least for this range of temperatures.

\bigskip\par

\vfill
\eject
\par
\pagebreak

\begin{table}
\center
\begin{tabular}{|c|c|c|c|}\hline
& & &\\ Author & Equation & $B(T)$ & $T_i$\\ & & &\\ \hline \hline & &
&\\ Van der Waals & $PV= RT + bP - {a\over V} + {ab\over V^2}$ & $
b-{a\over RT}$ & $\infty$\\ & & &\\ \hline & & &\\ Berthelot & $PV=RT
+ bP - {a\over TV} + {ab\over TV^2}$ & $b-{a\over RT^2}$ & $\infty$ \\
& & &\\ \hline & & &\\ Dieterici & $PV=RTexp(- {a\over RTV}) +bP$ & $
b-{a\over RT}$ & $\infty$ \\ & & &\\ \hline & & &\\ Redlich-Kwong &
$PV=RT + bP - {a\over \sqrt T}$ & $b$ & -- \\ & & &\\ \hline & & &\\
Beattie-Bridgeman & $PV=RT+{D\over V} +{E\over V^2} + {F\over V^3}$ &
${b - {a\over {RT}}} - {c\over T^3}$ & $\sqrt{{-3cR\over a}}$ \\ & &
&\\ \hline
\end{tabular}
\caption{Equations of state and respectives {\it B(T); a, b} and $c$
are different constants for each equation and gas while $D$, $E$ and
$F$ are functions of $T$ \protect{\cite{Cast}}. From these $B(T)$ and
solving $dB/dT = 0$ one finds $T_i$. The only possible choice for
finite $T_i$ among these equations is the Beattie-Bridgeman equation
of state. However, as $a$ and $c$ in this equation are positive
constants for all real gases, we can see that none of the above
equations lead to any real finite $T_i$.}
\end{table}

\begin{table}
\center
\begin{tabular}{|c||c|c|c|c||c|}\hline
Gas &$a$ x $10^5$ & $b$ x $10^7$ & $c$ x $10 ^3$ & $e$ x $10^{-1}$
&$T_i (K)$\\ \hline \hline Ar&251.00&-2.40&-972.00&-345.60&2015\\
\hline He&87.01&-3.31&-18.77&--&222\\ \hline
H$_2$&166.00&-7.66&-172.33&-- &474\\ \hline
N$_2$&357.46&-9.36&-1044.84&-242.53&1123\\ \hline
Ne&102.10&-0.80&-100.00&-31.33&1259\\ \hline
\end{tabular}
\caption{Coefficients of the eq.(\protect{\ref{HO}}) for $B(T)$ from
Holborn and Otto \protect{\cite{HO}} and $T_i$ calculated from
eq.(\protect{\ref{HO2}}).}
\end{table}

\begin{table}
\center
\begin{tabular}{|c|c|c|}\hline
j&$T_{i}^*$&$T^{*}_{iJT}$\\ \hline
\hline
1 & 19.69595658 & 4.289341659 \\ \hline
2 & 24.44023065 & 5.953282255 \\ \hline
3 & 25.06255633 & 6.323900009 \\ \hline
4 & 25.14189822 & 6.407536474 \\ \hline
5 & 25.15139334 & 6.425971440 \\ \hline
10 & 25.15257343 & 6.430797595 \\ \hline
15 & 25.15257343 & 6.430798467 \\ \hline
20 & 25.15257343 & 6.430798467 \\ \hline
\end{tabular}
\caption{Numerical results for the (reduced) inversion temperatures
$T_{i}^*$ and $T_{iJT}^*$ corresponding to the Lennard-Jones
potential.}
\end{table}

\begin{table}
\center
\begin{tabular}{|c|c|c|c|}\hline
Gas &$\epsilon_0/ k$&$ T_i (K)$&$ T_{iJT} (K)$\\ \hline
\hline
Ar&119.8&3013&770\\ \hline
He&10.22&257&65\\ \hline
Kr&171&4301&1100\\ \hline
H$_2$&37&930&238\\ \hline
N$_2$&95.05&2391&611\\ \hline
Ne&34.9&878&224\\ \hline
O$_2$&118&2968&759\\ \hline
Xe&221&5558&1421\\ \hline
\end{tabular}
\caption{Data for the parameter ratio $\epsilon_0/k$ of some real
gases for the Lennard-Jones potential from \protect{\cite{Hisch}} and
the corresponding inversion temperatures calculated through
eqs.(\protect{\ref{Temp1}}) and (\protect{\ref{Temp2}}).}
\end{table}

\begin{table}
\center
\begin{tabular}{|c|c|c|}\hline
Gas&$T_{iJT} (K) $&$T_i (K) $\\ \hline
\hline
Ar&723&2827\\ \hline
He&40&156\\ \hline
H$_2$&204&797\\ \hline
N$_2$&625&2444\\ \hline
Ne&270&1055\\ \hline
O$_2$&750&2933\\ \hline
\end{tabular}
\caption{The well known inversion temperatures $T_{iJT}$ for the
Joule-Thompson process from \protect{\cite{Kest}} and $T_i$ calculated
from them using eq.(\protect{\ref{rel}}).}
\end{table}

\begin{table}
\center
\begin{tabular}{|c||c|c|c|c|c||c|}\hline
Gas&$a$ x $10^5$& $b$ & $c$ & $d$ &$\chi^2$ & $T_i (K)$\\ \hline
\hline
Ar&-3.8458&1.6470&1.4676&-0.42510&39.84&-\\ \hline
He&-0.00037874&0.19433&-731.41&1.1205&7.80&162\\ \hline
Kr&12.884&1.7316&0.2639&-0.68291&7.36&-\\ \hline
H$_2$&0.053319&1.1250&62.656&0.1747&4.77&762\\ \hline
N$_2$&-1.7851&1.4891&-0.075169&21.124&8.072&-\\ \hline
Ne&0.0009038&0.24146&-3873.2&1.0229&0.99&-\\ \hline
O$_2$&-12.758&1.9062&7.3866(-4)&-1.6421&3.20&-\\ \hline
Xe&-5.8075&1.4302&20.145&-0.11510&54.04&-\\ \hline
\end{tabular}
\caption{Coefficients of eq.(\protect{\ref{b1}}) fitting experimental
data compiled in \protect{\cite{DS}} and the corresponding $T_i$. See
captions of figs. 2 -- 9 for details concerning experimental data used
in the numerical treatment for each gas.}
\end{table}

{\footnotesize
\begin{table}
\center
\begin{tabular}{|c||c|c|c|c|c|c|c||c|}\hline
Gas& $a$ & $b$ & $c$ & $d$ & $e$ & $f$ &$\chi^2$ &$T_i (K)$\\ \hline \hline
Ar&-1.3342(5)&1.4215&-3.2404(8)&3.6778&11.018&-0.14186&33.10&-\\
\hline He&-3244.5&1.8348&5.5676&-0.29099&-0.79394&-0.54227&3.81&225\\
\hline
Kr&6.9839(49)&20.404&3.0730(6)&1.8976&0.062216&-0.88609&6.78&-\\
\hline H$_2$&-10127&1.0608&5074.5&1.0204&65.673&0.18220&4.76&742\\
\hline
N$_2$&-27.083&1.0338&-3.0304(10)&4.4904&368.82&0.29037&6.59&1779\\
\hline Ne&-3203.8&0.85450&4.8903(-4)&-1.1751&426.34&0.43549&0.91&-\\
\hline
O$_2$&7.7283(-4)&-1.6345&-1.2825(6)&1.9066&25352&2.2728&3.20&-\\
\hline Xe&8.2799(5)&1.2173&-1.3374(6)&1.2710&2.9953&-0.36180&53.67&-\\
\hline
\end{tabular}
\caption{Coefficients of eq.(\protect{\ref{b2}}) fitting experimental
data compiled in \protect{\cite{DS}} and the corresponding $T_i$. The
number between parenthesis represents the decimal power of each
coefficient.  See captions of figs. 2 -- 9 for details concerning
experimental data used in the numerical treatment for each gas.}
\end{table}
}

{\footnotesize
\begin{table}
\center
\begin{tabular}{|c||c|c|c|c|c|c|c||c|}\hline
Gas& $a$ & $b$ & $c$ & $d$ & $f$ & $g$ & $\chi^2$ & $T_i (K)$\\ \hline
\hline Ar&-1.7016&1.4586&-6.6200(15)&7.6571&24.387&-1.8960&28.21&-\\
\hline He&-608.29&1.1845&-2.1380(14)&11.665&14.480&5.2356&2.14&199\\
\hline Kr&-1.0322(7)&2.0942&1.5424(16)&6.2812&7.7358&-13.232&6.64&-\\
\hline
H$_2$&-1.7658(5)&1.2639&1.6979(5)&1.2655&22.693&3.1394&4.58&574\\
\hline N$_2$&42.198&0.050811&-1.5131(6)&2.0026&-73.672&44.518&6.32&-\\
\hline Ne&-7354.0&1.2176&43.256&0.14858&2017.9&1142.3&0.90&873\\
\hline
O$_2$&-6.6122(5)&1.7889&-1.5766(6)&2.3994&3.5171&-37.111&3.07&-\\
\hline
Xe&-5.6145(6)&1.7824&1.7444(8)&2.6786&23.081&4.5891&52.82&1870\\
\hline
\end{tabular}
\caption{Coefficients of eq.(\protect{\ref{b3}}) fitting experimental
data compiled in \protect{\cite{DS}} and the corresponding $T_i$. The
number between parenthesis represents the decimal power of each
coefficient.  See captions of figs. 2 -- 9 for details concerning
experimental data used in the numerical treatment for each gas.}
\end{table}
}

{\tiny
\begin{table}
\center
\begin{tabular}{|c||c|c|c|c|c|c|c|c|c||c|}\hline
Gas& $a$ & $b$ & $c$ & $d$ & $e$ & $f$ & $g$ & $h$ & $
\chi^2$&$T_i (K)$\\ \hline \hline
Ar&-40619&1.0683&-3.7726(15)&7.4168&154.41
&0.17160&77.936&48.300&26.47&3841\\
\hline
He&-4.5828(-3)&-0.50182&-606.47&1.1823
&-6.6369(15)&12.931&14.540&5.1296&2.12&199\\
\hline
Kr&-1.8199(6)&1.8400&-3.3257(8)&2.8883
&1.0915(13)&4.8686&9.1254&-11.701&6.58&-\\
\hline
H$_2$&-123.49&1.1909&5960.4&1.1763&19.971
&0.86958&23.248&3.5224&4.58&549\\
\hline
N$_2$&-14734&0.95513&-3.1691(7)&2.8771
&155.79&0.16280&-9.3920(-3)&-69.007&6.24&3130\\
\hline
Ne&-12337&1.3439&2.4123&-0.053448&34.092
&0.15637&-128.42&565.23&0.88&-\\
\hline
O$_2$&4.9441(-7)&-2.8490&-3.5770(6)&2.1115
&3.2702(14)&6.5228&-0.025554&-85.016&1.90&-\\
\hline
Xe&-1.8835(5)&1.0629&2.4642(7)&2.3383&239.78
&0.10465&326.00&28.401&52.25&-\\
\hline
\end{tabular}
\caption{Coefficients of eq.(\protect{\ref{b4}}) fitting experimental
data compiled in \protect{\cite{DS}} and the corresponding $T_i$. The
number between parenthesis represents the decimal power of each
coefficient.  See captions of figs. 2 -- 9 for details concerning
experimental data used in the numerical treatment for each gas.}
\end{table}
}

\begin{table}
\center
\begin{tabular}{|c|c|}\hline
Gas& $T_i (K)$ \\ \hline
\hline
Ar&$ (3.8 \pm 1.8) 10^3$\\ \hline
He&$ (2.0 \pm 0.5) 10^2$\\ \hline
Kr&-\\ \hline
H$_2$&$ (6.6 \pm 2.7) 10^2$ \\ \hline
N$_2$&$ (2.5 \pm 1.4) 10^3$ \\ \hline
Ne&$ (8.7 \pm 3.9) 10^2$ \\ \hline
O$_2$&-\\ \hline
Xe&$ (1.9 \pm 0.5) 10^3$ \\ \hline
\end{tabular}
\caption{Inversion temperatures $T_i$ calculated from the numerical
analysis presented in tables 6 to 9, using a $\chi^2$ weighted
average. The uncertainties were calculated taken into account $T_i$
data from all tables 2, 4 -- 9. The Krypton and Oxygen do not present
inversion temperatures in this table since the experimental data
available does not permit us to find them (see also fig. 10).}
\end{table}

\end{document}